\documentclass[english,aps, paper, twocolumn]{revtex4}
\usepackage[T1]{fontenc}
\usepackage[utf8]{inputenc}
\usepackage{babel}
\usepackage{amssymb}
\usepackage{graphicx}
\usepackage{esint}
\usepackage[unicode=true,pdfusetitle,
 bookmarks=true,bookmarksnumbered=false,bookmarksopen=false,
 breaklinks=false,pdfborder={0 0 1},backref=false,colorlinks=false]
 {hyperref}

\makeatletter
\@ifundefined{textcolor}{}
{%
 \definecolor{BLACK}{gray}{0}
 \definecolor{WHITE}{gray}{1}
 \definecolor{RED}{rgb}{1,0,0}
 \definecolor{GREEN}{rgb}{0,1,0}
 \definecolor{BLUE}{rgb}{0,0,1}
 \definecolor{CYAN}{cmyk}{1,0,0,0}
 \definecolor{MAGENTA}{cmyk}{0,1,0,0}
 \definecolor{YELLOW}{cmyk}{0,0,1,0}
}

\usepackage{lmodern}  

\makeatother

\begin{document}
\global\long\def\vr{\mathbf{r}}
\global\long\def\vrp{\mathbf{r'}}
\global\long\def\o{\omega}
\global\long\def\vk{\mathbf{k}}
\global\long\def\vkp{\mathbf{k'}}

\title{Rayleigh scattering in coupled microcavities: Theory}

\author{Z. Vörös, G. Weihs}

\affiliation{Department of Experimental Physics, University of Innsbruck, Technikerstraße
25/d, Innsbruck, A-6020 Austria}
\begin{abstract}
In this paper we theoretically study how structural disorder in coupled
semiconductor heterostructures influences single-particle scattering
events that would otherwise be forbidden by symmetry. We extend the
model of V. Savona \cite{Savona2007} to describe Rayleigh scattering
in coupled planar microcavity structures, and answer the question,
whether effective filter theories can be ruled out. They can.
\end{abstract}
\maketitle

\section{Introduction}

Resonant Rayleigh scattering (RRS) caused by the structural disorder
of semiconductor heterostructures is a well-known phenomenon in semiconductor
optics. It is usually regarded as an unwanted feature, however, it
has recently been realised that it can be exploited to study fundamental
effects, e.g., to probe for polariton superfluidity \cite{Carusotto2004},
or measure the properties of a polariton condensate \cite{Christmann2012}.
Recently, there have also been attempts at influencing RRS by inducing
strain in the heterostructures \cite{Zajac2012c}, or by superposing
a periodic potential on the random disorder \cite{Abbarchi2012}. 

In the early stages of RRS experiments, when the phenomenon was first
observed in polaritonic systems, it was not clear, what exactly causes
Rayleigh scattering: since polaritons are a coherent superposition
of quantum well excitons and cavity photons, in principle, either
of the constituents could be affected by structural disorder. However,
given that RRS did not show a strong temperature dependence, and that
it could be observed even in cavities without quantum well excitons
(so-call cold cavities), it was soon established that photonic imperfections
of the structures are responsible \cite{Houdre2000,Houdre2001,Gurioli2001,Gurioli2002,Gurioli2003,Langbein2004,Langbein2010,Zajac2012}. 

Due to the resonant nature of the scattering, in single microcavities,
the final states will be distributed on a circle, that can easily
be distinguished in the far-field emission pattern. It is important
to realize that in such a case, scattering happens on a single branch,
given that the lower and upper polaritons are energetically separated.
However, when one couples two such cavities through partially reflecting
mirrors, the resulting eigenmodes of the system overlap energetically.
It is only natural to ask, whether RRS still happens only on the branch
that is pumped (intra-branch scattering), and if this is not the case,
what determines the conditions for scattering to the unpumped branch
(inter-branch scattering). What makes the problem even more interesting,
is the fact that in a perfect structure (i.e., one without disorder),
inter-branch scattering would be forbidden by the reflection symmetry
(parity) of the system, and one could then ask, to what extent disorder
breaks this symmetry. The answers to these questions have also some
practical consequences: some of the parametric scattering schemes
that attempt to generate quantum correlated light are degenerate,
and it is of paramount importance to find out, whether parametric
luminescence has to compete with RRS, which would reduce the quantum
nature of the generated light.

However, even after it was established that only photons undergo Rayleigh
scattering, there could still be various ways in which scattering
could happen. The simplest explanation is given by the so-called filter
theory, which states that the intra-cavity light field is scattered
by the disorder \emph{inside} the cavity producing a spherical wave,
and only those components of the scattered field can be detected outside
that are resonant with the cavity, for the other components will be
attenuated. In this regard, symmetry plays no role in the scattering
process, and the situation is identical to that of a Fabry-Pérot resonator
with a diffusive plate between the mirrors \cite{Freixanet1999}.
In particular, since the cavity filters the modes of the spherical
wave, it is assumed in this model that the cavity mirrors are perfect. 

A cavity consists of two mirrors, and the enclosed space. The filter
theory assumes that scattering happens in the enclosed space. But
RRS could also be caused by the imperfections of the mirrors. This
is the starting point of the theory of V. Savona \cite{Savona2007,Sarchi2009},
in which he proves rigorously that because of the large effective
mass of the exciton (about 5 orders of magnitude larger than that
of the photons), the effect of excitonic disorder is negligible in
RRS. While his theory accounts for all the observed phenomena, it
still cannot unambiguously rule out the filter theory. In this contribution,
we extend his theory to multiple cavities, and we analyse the case
of double cavities in detail. For such structures, the filter theory
would predict that the scattered spherical wave would be filtered
by two resonances, resulting in the far field in two concentric rings
with approximately equal powers. Savona's theory, on the other hand,
predicts that the far field emission will contain the same two concentric
rings as in the filter theory, but their relative intensities will
depend on the correlations of the photonic disorder. It can thus be
expected that a measurement on the relative intensities of the two
rings would differentiate between the two models.

In the first half of the paper, in Section II., we outline the underlying
principles of Savona's theory of Rayleigh scattering in a single microcavity.
In Section III., we introduce simplifications that do not change the
main conclusions, and extend the model to the case of coupled cavities.
We solve the simplified equations numerically, and show that in coupled
cavities, \emph{longitudinal} correlations of the the disorder potential
are the relevant parameter of RRS. While this model is conceptually
simple, it can only be solved numerically. In the second part of the
paper, starting with Section IV., we apply perturbation theory, and
develop a simple analytical model that can reproduce most of the results
of the numerical simulations. In the appendix, we show how this model
can be extended to triple cavities, a rather popular experimental
realization of coupled cavities. We conclude with a short conclusion
and outlook.

\section{Rayleigh scattering in a single microcavity}

First we re-capitulate Savona's theory of for the case of a single
cavity. The extension of the model to an arbitrary number of cavities
is straightforward. The theory is based on the assumption that the
electric field inside the cavity can be written as 

\[
E(\vr,z)=E(\vr)\exp(ik_{z}(\vr)z)\,,
\]
where $k_{z}(\vr)$ is the position-dependent photon momentum along
the $z$ direction, and $\vr$ lies in the plane of the cavities.
Above, only a single mode is considered, therefore, the polarization
degree of freedom plays no role. By introducing the macroscopic exciton
polarization field $P(\vr)$, the Helmholtz equation 

\[
\nabla_{\vr}^{2}E(\vr)+\left(\frac{\omega_{p}}{c^{2}}\epsilon_{0}-k_{z}^{2}(\vr)\right)E(\vr)+4\pi\frac{\omega^{2}}{c^{2}}P(\vr)=0
\]
can then be expanded around the mean cavity length, $\lambda_{0}$,
resulting in an effective Schrödinger equation for the photon mode

\begin{eqnarray}
i\hbar\frac{\partial E(\vr,z,t)}{\partial t} & = & -\frac{\hbar^{2}}{2m_{p}}\nabla_{\vr}^{2}E(\vr,t)+V_{p}(\vr)E(\vr,t)\nonumber \\
 &  & \qquad-2\pi\frac{\hbar\omega_{p}}{\epsilon_{0}}P(\vr,t)\label{eq:linearised_maxwell}
\end{eqnarray}
where the effective mass of the photon is given by the expressions
\[
m_{p}=\frac{\hbar\omega_{p}}{2c^{2}}\qquad\mathrm{and}\qquad\omega_{p}=\frac{2\pi c}{n\left(\lambda_{0}+L_{\mathrm{DBR}}\right)}\,,
\]
where $L_{\mathrm{DBR}}$ is the penetration length of the electric
field into the distributed Bragg reflector (DBR), and $n$ is the
refractive index of the cavity. The disorder potential, $V_{p}(\vr)$,
stems from the local variations of the cavity length, $\delta L(\vr)$,
as 

\[
V_{p}(\vr)=\frac{\hbar c}{\sqrt{\epsilon_{0}}}\delta k_{z}(\vr)=-\hbar\omega_{p}\frac{\delta L(\vr)}{\lambda_{0}+L_{\mathrm{DBR}}}\,.
\]
Eq.\,(\ref{eq:linearised_maxwell}) can be completed by noticing
that the polarization, $P(\vr)$, depends linearly on the center-of-mass
wave function $\Psi(\vr,t)$ of the exciton \cite{Savona2000}, whence
one arrives at the coupled equations

\begin{widetext}

\begin{eqnarray}
i\hbar\frac{\partial E(\vr,t)}{\partial t} & = & \hbar\omega_{p}(\vr)E(\vr,t)-\frac{\hbar^{2}}{2m_{p}}\nabla_{\vr}^{2}E(\vr,t)+V_{p}(\vr)E(\vr,t)+\frac{\hbar\Omega_{R}}{2}\Psi(\vr,t)+f(\vr,t)\label{eq:single_cavity_E}\\
i\hbar\frac{\partial\Psi(\vr,t)}{\partial t} & = & \hbar\omega_{x}(\vr)E(\vr,t)-\frac{\hbar^{2}}{2m_{x}}\nabla_{\vr}^{2}\Psi(\vr,t)+V_{x}(\vr)\Psi(\vr,t)+\frac{\hbar\Omega_{R}}{2}E(\vr,t)\,,\label{eq:single_cavity_X}
\end{eqnarray}

\end{widetext}

Eqs.\ (\ref{eq:single_cavity_E}-\ref{eq:single_cavity_X}) constitute
the equation of motion of the coupled photon-exciton system. All quantities
with the subscript $x$ refer to the exciton component, while the
subscript $p$ stands for the cavity photon. $f(\vr,t)$ is a source
term that generates intracavity photons (and through the coupling,
excitons), while $\omega_{p}(\vr)$, and $\omega_{x}(\vr)$ contain
the photonic and excitonic disorder, and the finite lifetime. Consequently,
the frequencies can have imaginary parts. Without disorder, these
equations can be diagonalized, resulting in two new eigenmodes, the
so-called upper and lower polaritons, whose splitting in energy is
usually called the Rabi splitting, $\hbar\Omega_{R}$.

When the disorder is not zero, due to the large mis-match of the exciton
and photon effective masses, the numerical solution of these two equations
is not trivial, and requires huge resources. However, a key observation
that can be drawn from Savona's calculations is that Rayleigh scattering
is dominated by the photonic disorder, and the sole role of the quantum
well exciton is to modify the dispersion relations of the polariton
states. This statement is also confirmed by the experimental results
of Maragkou et al., where Rayleigh scattering was observed even in
a cold cavity \cite{Maragkou2011}. For this reason, we will drop
the equation for the exciton wavefunction. In cases, where accurate
modeling of the polariton states is required, in order to reduce the
computation costs, one can introduce higher-order derivatives in Eq.~(\ref{eq:single_cavity_E}),
so as to better approximate the non-parabolicity of the dispersion.
However, since this would not qualitatively change our results, we
will not pursue this route here.

\section{Coupled cavities}

\subsection{Model equations}

Having discussed the case of a single cavity, we now turn to the derivation
of the equation of motion in coupled microcavities. 

\begin{figure}[h]
\includegraphics[width=0.78\columnwidth]{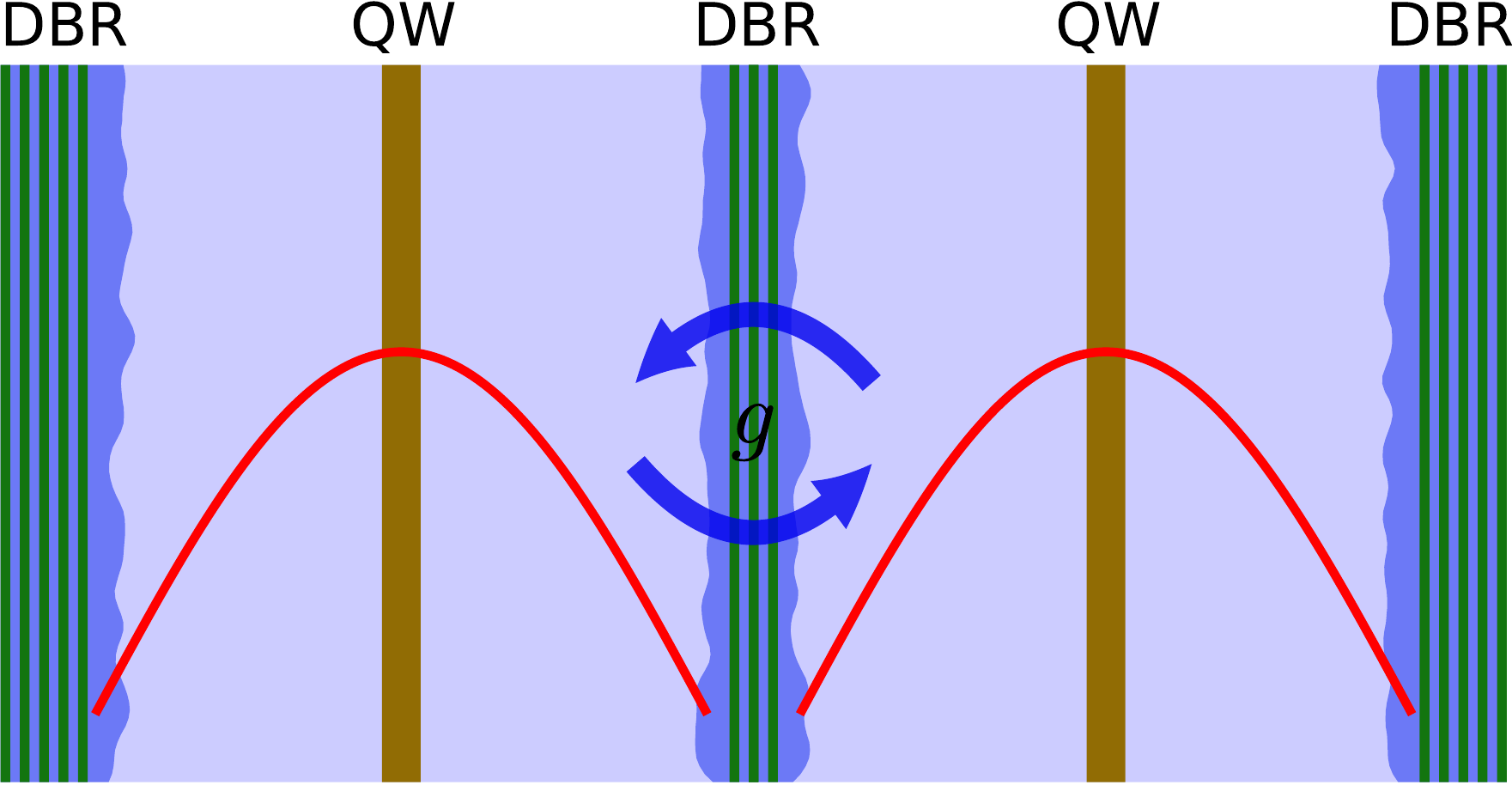}\includegraphics[width=0.21\columnwidth]{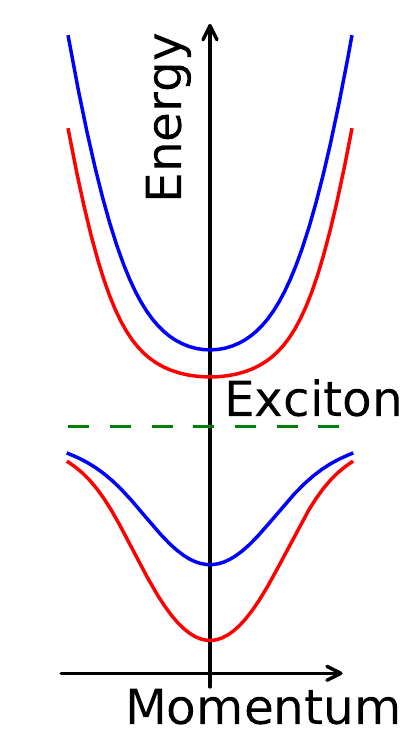}

\caption{Left: The system under consideration. The disorder potential acting
on the photonic modes is the rugged mirror surface. The two uncoupled
modes are the solid red lines, while the coupling of the cavities,
$g$, is symbolized by the blue arrows. Right: schematic polariton
dispersions in a coupled cavity at zero detuning. Also shown is the
exciton line in dashed green.}

\label{fig:disordered_cavity}
\end{figure}

The system under consideration is depicted in Fig.\,\ref{fig:disordered_cavity}:
two cavities are coupled through a partially transparent DBR. Quantum
wells are located in the center of the cavities. The disorder of the
mirrors is symbolized by their rugged surface. 

Out of the two longitudinally localized photonic modes (red solid
lines in Fig.\,\ref{fig:disordered_cavity}), the coupling produces
four polariton eigenstates that are split. Just as in the single cavity
case, the lower polaritons are still energetically separated from
the upper polaritons, but both the upper, and the lower polaritons
overlap spectrally. This raises the possibility of scattering processes
in which no energy is exchanged, but which take a polariton from one
branch to another (inter-branch scattering), as shown on the right
hand side of Fig.\,\ref{fig:disordered_cavity}. While in a perfect
cavity such processes would be forbidden by the conservation of parity,
for a disordered system parity is not a good quantum number anymore.
In the rest of the paper, we will study how inter-branch scattering
depends on the properties of the disorder that breaks the parity symmetry.

As mentioned above, in order to simplify the description, we drop
the excitonic component. We note, nevertheless, that in coupled cavities,
one can still retain the excitonic wavefunction, but the coupling
will be mediated by the photonic part only: excitons are confined
to their respective quantum wells, and in realistic experimental situations,
exciton tunneling between two quantum wells can safely be ignored
\cite{Einkemmer2013}.

In what follows, we will denote the transverse wavefunction of the
photon fields in the two cavities by $E_{a,b}(\vr,t)$, the corresponding
energies by $\o_{a,b}(\vr)$, the photon effective masses by $m_{a,b}$,
and the source terms by $f_{a,b}$. 

Moreover, we denote the coupling of the two cavities by $g_{0}$.
In general, the transmission of the DBRs depends on the wavevector,
and by applying the standard transfer matrix approach, it can be shown
that this dependence results in a correction quadratic in the length
of the wavevector. In the momentum space equations, this correction
can be accounted for by adding the Laplace operator to the coupling
constant as $\epsilon\nabla^{2}$. However, in subsequent calculations,
we drop this term, for it only modifies the dispersion relations,
but has no other bearing on our results. 

The energies $\o_{a,b}(\vr)$ depend, on the one hand, on the length
of the cavities (as do the effective masses $m_{a,b}$), and on the
other hand, on the disorder potential. It is because of this spatially
inhomogeneous disorder potential that the usual Bogolyubov transformation
cannot be carried out, and that we cannot decompose the eigenmodes
of the coupled system as a symmetric and an anti-symmetric linear
combination of the original states. 

With these considerations, the two equations governing the time-evolution
of the two modes in question are 

\begin{widetext}

\begin{eqnarray}
\frac{\partial E_{a}(\vr,t)}{\partial t} & = & i\o_{a}(\vr)E_{a}(\vr,t)+\frac{i}{m_{a}}\nabla^{2}E_{a}(\vr,t)+i(g_{0}-\epsilon\nabla^{2})E_{b}(\vr,t)+f_{a}(\vr,\omega_{0},t)\label{eq:mode_a}\\
\frac{\partial E_{b}(\vr,t)}{\partial t} & = & i\o_{b}(\vr)E_{b}(\vr,t)+\frac{i}{m_{b}}\nabla^{2}E_{b}(\vr,t)+i(g_{0}-\epsilon\nabla^{2})E_{a}(\vr,t)+f_{b}(\vr,\omega_{0},t)\,.\label{eq:mode_b}
\end{eqnarray}
\end{widetext}

In the following, we will assume that the two cavities have the same
length, and a disorder potential with the same statistical properties.
This also implies $m_{a}=m_{b}$, and $\o_{a}(\vr)=\o_{0}+V_{a}(\vr)$,
and $\o_{b}(\vr)=\o_{0}+V_{b}(\vr)$, where $V_{a,b}(\vr)$ are the
disorder potentials with zero mean, and $\langle V_{a}(\vr)V_{a}(\vrp)\rangle_{\mathbf{r}}=\langle V_{b}(\vr)V_{b}(\vrp)\rangle_{\mathbf{r}}=V_{0}^{2}e^{-\vert\vr-\vrp\vert^{2}/\xi^{2}}$,
where $\xi$ is the \emph{transverse} correlation length, and $\langle...\rangle_{\vr}$
stands for averaging over the spatial coordinate. As mentioned above,
we also set $\epsilon=0$.  

Eqs.(\ref{eq:mode_a}-\ref{eq:mode_b}) describe two sets of harmonic
oscillators, which are pairwise coupled through the term $g_{0}-\epsilon\nabla^{2}$,
and where both sets have some inhomogeneous broadening given by $V_{a,b}(\vr)$.
In general, there is no reason, why $V_{a}(\vr)$, and $V_{b}(\vr)$
should be correlated in any way. However, semiconductor heterostructures
are special in the sense that the consecutive layers deposited on
the substrate are produced by the same device. Therefore, if the spatial
distribution of the deposited material does not change in time, then
it is only natural to expect the successive layers to display some
statistical correlation. Conversely, the longitudinal correlations
are a measure of the long-term stability of the growth process.

\subsection{Numerical results}

We have already stipulated some statistical properties of the disorder
potentials; in order to characterize their independence, we define
their transverse cross-correlation as 

\begin{equation}
\mathcal{C}=\frac{\langle V_{a}(\vr)V_{b}(\vr)\rangle_{\vr}}{\sqrt{\langle\vert V_{a}(\vr)\vert^{2}\rangle_{\vr}\langle\vert V_{b}(\vr)\vert^{2}\rangle_{\vr}}}\,.
\end{equation}
For our numerical simulations, these potentials will be generated
as 

\begin{eqnarray}
V_{a}(\vr) & = & V_{0}\int\, d\vrp K(\vr,\vrp,\xi)N_{a}(\vrp)\label{eq:va_numerical}\\
V_{b}(\vr) & = & \mathcal{C}V_{a}(\vr)+V_{0}\sqrt{1-\mathcal{C}^{2}}\int\, d\vrp K(\vr,\vrp,\xi)N_{b}(\vrp)\label{eq:vb_numerical}
\end{eqnarray}
where $N_{a,b}(\vr)$ are two independent normally distributed random
variables with zero mean, and width $1$, and $K(\vr,\vrp,\xi)$ is
a Gaussian kernel, which introduces the transverse correlations. Since
the distributions are centered on zero, $V_{0}$ gives the average
of the potential fluctuations. Moreover, the particular construction
in Eq.\,(\ref{eq:vb_numerical}) ensures that $V_{b}(\vr)$ is also
a normal distribution. 

The equations in Eqs.\,(\ref{eq:mode_a}-\ref{eq:mode_b}) are solved
numerically in conjunction with Eqs.\,(\ref{eq:va_numerical}-\ref{eq:vb_numerical})
on a grid of 512 by 512 points with a spatial extent of 200 by 200
$\mu$m over a time domain of 5 ps with 800 points. In the spatial
coordinates, we apply periodic boundary conditions, which slightly
modify the exact shape of the dispersion, namely, by definition, the
dispersion will have zero slope at the zone boundaries. However, this
does not change any of our conclusions. 

The source terms obey the relation $f_{b}(\vr,\omega_{0},t)=\pm f_{a}(\vr,\omega_{0},t)$,
depending on whether we are trying to drive the symmetric, or anti-symmetric
mode of the \emph{unperturbed} ($\o_{a,b}(\vr)=\mathrm{{constant}}$)
system. In the subsequent discussion, whenever we refer to the symmetric
or anti-symmetric modes, we mean the modes that emerge in the case
$V_{a}(\vr)=V_{b}(\vr)=0$. The source terms $f_{a,b}(\vr,\omega_{0},t)=\pm\exp(i\omega_{0}t)\exp(-(t-t_{0})^{2}/\tau^{2})\exp(-|\vr|^{2}/\sigma^{2})$
are assumed to be short Gaussian pulses with duration $\tau=1$\,ps,
spatial extent $\sigma=30\,\mu$m, and a carrier frequency $\omega_{0}$
equalling the resonance energy of the corresponding polariton branch.
In order to simulate a finite initial momentum $\vk_{0}$, $f_{a,b}(\vr,\omega,t)$
are multiplied by a phase term $\exp(-i\vk_{0}\cdot\vr)$. 

\begin{figure}[h]
\includegraphics[width=1\columnwidth]{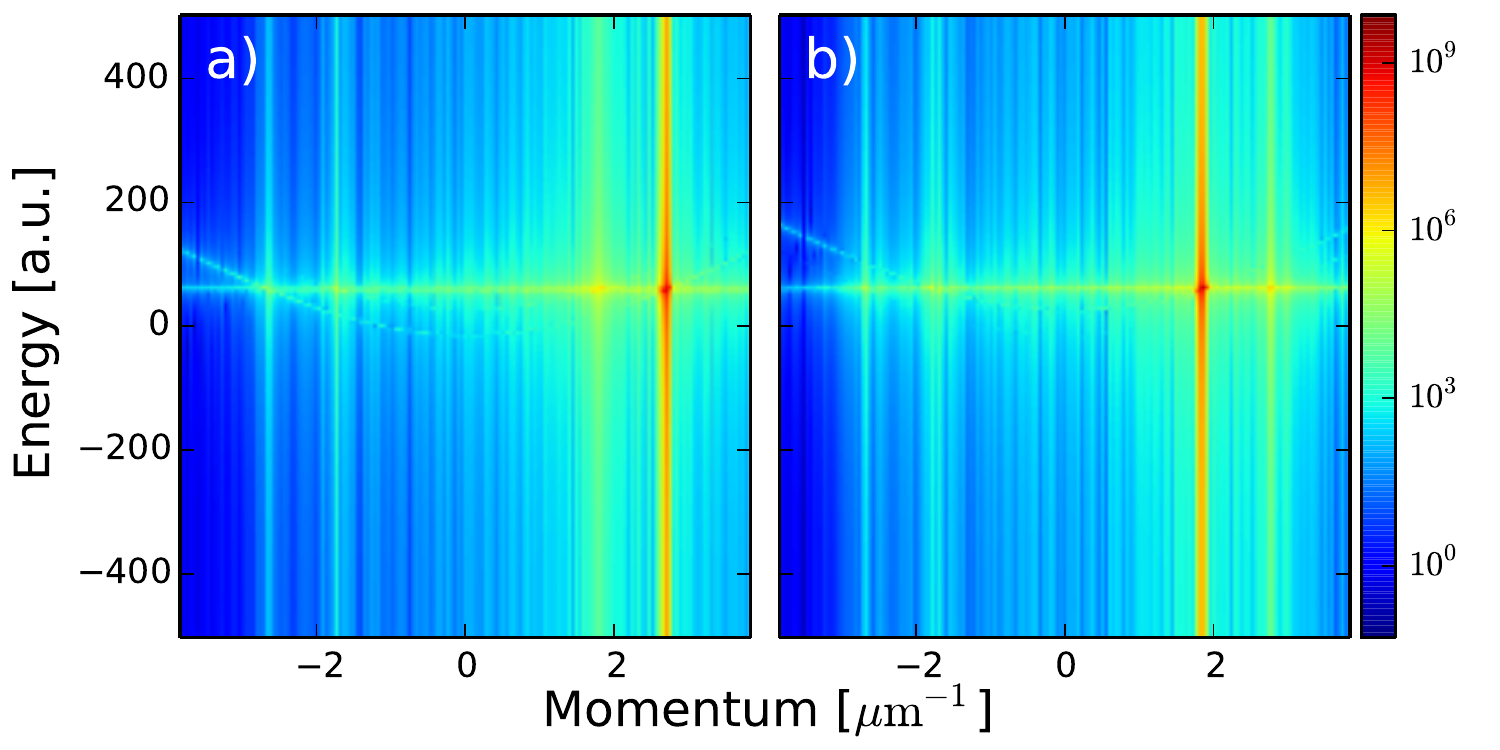}

\caption{Logarithmic plot of the Rayleigh scattering spectrum of the two photon
modes. The sample is excited by a spectrally narrow resonant laser
pulse on either the symmetric (a), or anti-symmetric (b) branch. The
correlation is $\mathcal{C}=0$, and the disorder potential is $V_{0}=0.1\,$meV,
while the correlation length is $\xi=0.5\,\mu$m.}

\label{fig:dispersion}
\end{figure}

In Fig.\,\ref{fig:dispersion} we plot the resulting spectrum as
a function of the in-plane momentum for the case when the disorder
correlation is $\mathcal{C}=0$. We excite the sample on either the
symmetric (a), or the anti-symmetric mode (b). As pointed out above,
the strictly parabolic dispersion that one would expect from the Laplace
operator is modified by the periodic boundary conditions, and this
can be seen at higher momenta as a decrease in the slope of the dispersion
curves.

It is immediataly clear from this figure that the photon fields acquire
a non-zero amplitude even on the unpumped branches, and that this
happens irrespective of the symmetry of the pumped branch. 

\begin{figure}[h]
\includegraphics[width=1\columnwidth]{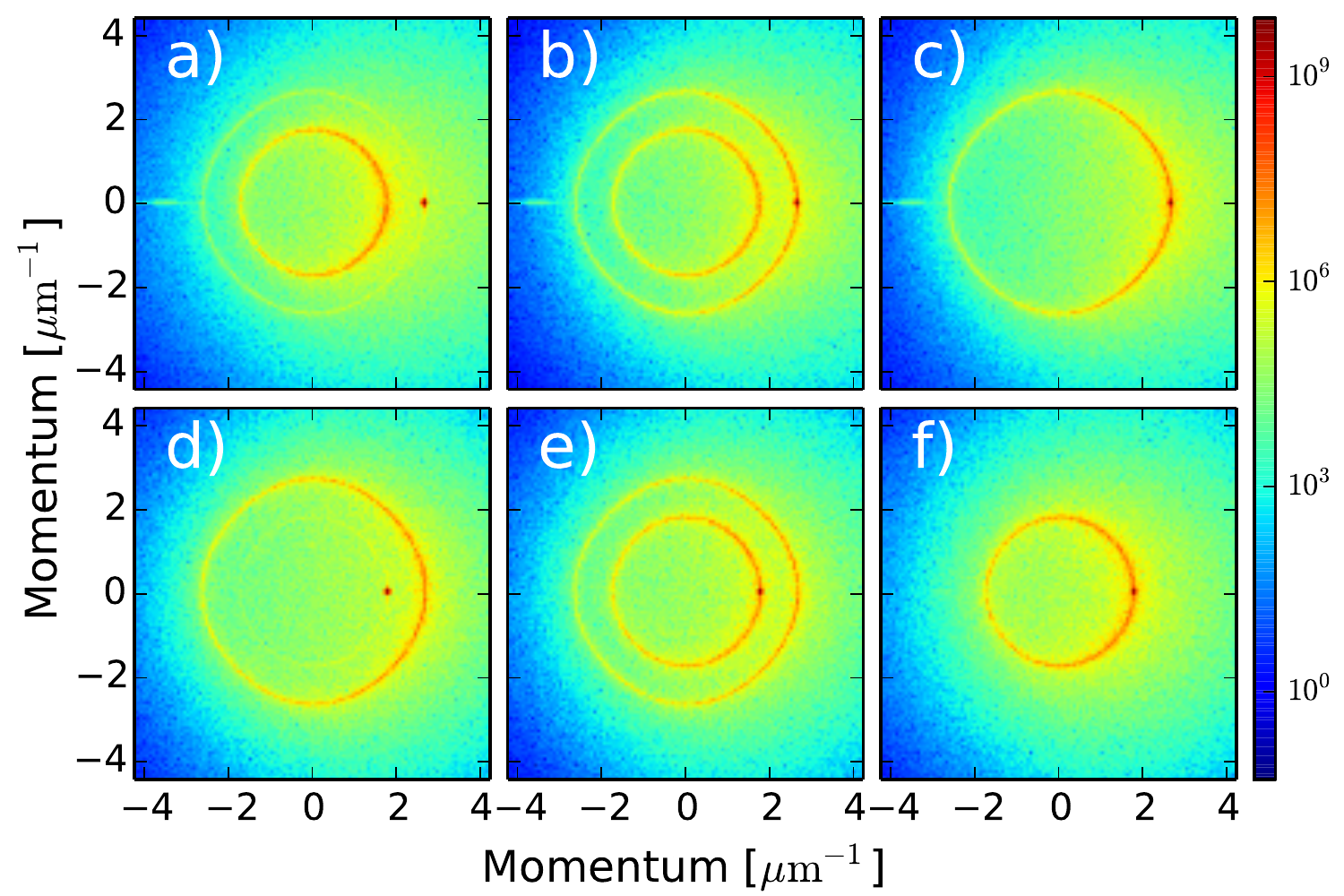}

\caption{Logarithmic plot of the far-field emission power. The sample is excited
by a spectrally narrow resonant laser pulse on either the anti-symmetric
(a-c), or the symmetric (d-f) mode. The excitation momentum can be
recognized as an intensity maximum at $\vk=(2.8,0)$, and $\vk=(1.9,0)\,\mu\mathrm{m}^{-1}$,
respectively. The correlations of the disorder potential are $\mathcal{C}=-1$
(a, d), $\mathcal{C}=0$ (b, e), and $\mathcal{C}=1$ (c, f), while
the transverse correlation length is $\xi=0.5\,\mu\mathrm{m}^{-1}$}

\label{fig:far-field1}
\end{figure}

In Fig.\ref{fig:far-field1}, we show the calculated far-field luminescence
for three values of the disorder correlation (totally anti-correlated,
$\mathcal{C}=-1$, uncorrelated, $\mathcal{C}=0$, and correlated
$\mathcal{C}=1$), and for two different values of the pump momentum.
In all six cases the pump is resonant with one of the polariton branches.
Assuming cylindrically symmetric dispersions (isotropic effective
masses), the Rayleigh rings should also be circular. The fact that
they are skewed in the figures is a consequence of the periodic boundary
conditions that we applied, but this will affect none of our conclusions. 

There are a couple of general observations that we can make at this
point. First, in all cases, the scattered intensity drops as a function
of the distance from the pump position. This is a consequence of the
finite transverse correlation length, and it has been observed experimentally,
e.g., in Houdré et al. \cite{Houdre2000}.

\begin{figure}[h]
\includegraphics[width=1\columnwidth]{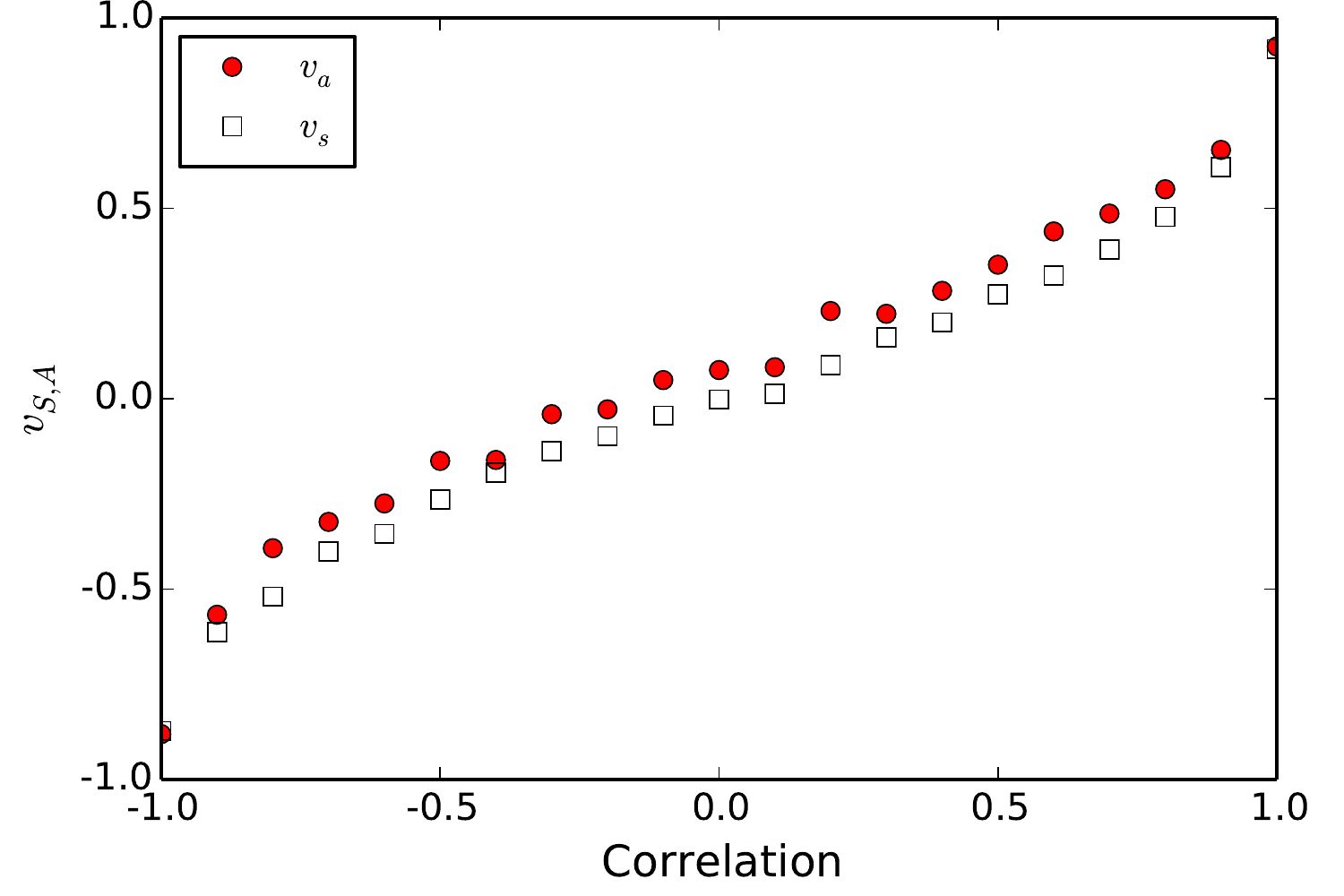}

\caption{The dependence of the inter-branch scattering intensitiy on the transverse
cross-correlation. The simulations parameters are the same as in Fig.\,\ref{fig:far-field1},
except for the correlation $\mathcal{C}$. The intensities are integrated
over circular shells of radii $1.85-1.95$, and $2.75-2.85\,\mu\mathrm{m}^{-1}$,
respectively. The transverse correlation length is $\xi=0.5\,\mu\mathrm{m}^{-1}$.}

\label{fig:correlation_dependence}
\end{figure}

Second, for the totally anti-correlated disorder ($\mathcal{C}=-1$),
intra-branch scattering is suppressed, and inter-branch scattering
is maximum. The opposite is true for the case of totally correlated
disorder ($\mathcal{C}=1$), while for uncorrelated disorder ($\mathcal{C}=0$),
both inter-branch, and inter-branch scattering happens. This general
trend is demonstrated more quantitatively in Fig.\,\ref{fig:correlation_dependence},
where we plotted the amount of inter-, and intra-branch scattering
as a function of the correlation, $\mathcal{C}$. As a measure of
inter-branch scattering, we define the visibility-like quantities

\begin{equation}
v_{S}=\frac{{\displaystyle \int_{SS}}\, I(\vk)-{\displaystyle \int_{SA}}\, I(\vk)}{{\displaystyle \int_{SS}}\, I(\vk)+{\displaystyle \int_{SA}}\, I(\vk)}\,,\quad v_{A}=\frac{{\displaystyle \int_{AA}}\, I(\vk)-{\displaystyle \int_{AS}}\, I(\vk)}{{\displaystyle \int_{AA}}\, I(\vk)+{\displaystyle \int_{AS}}\, I(\vk)}\label{eq:visibilities}
\end{equation}
where the integration is over the pumped, or unpumped polariton branch.
In the integrals, the first subscript designates the pumped branch,
while the second subscript denotes the integration contour's branch.
Instead of using the ratio of the integrated intensities, the particular
definition of $v_{A,B}$ is still normalized, but is devoid of the
singularities that we would encounter when the scattering intensity
tends to zero. Note that at this point, we did not stipulate the integration
domain: depending on the experimental configuration, it could be the
whole resonant circle, or, e.g., only half of it. (Experimentally,
not the whole circle might be accessible, because the reflected light
is usually masked.) The exact value of $v_{S,A}$ will, of course,
depend on the integration domain, but the general trends will be unaffected.
In order to account for the finite linewidth of the luminescence,
and to average out statistical fluctuations, we integrate over a narrow
circular shell of angular extent $2\pi$ and of radii $1.85-1.95$,
and $2.75-2.85\,\mu\mathrm{m}^{-1}$, respectively. The pump region
was excluded from the integration domain, so as to consider the scattered
intensities only. 

\section{Analytical model}

In this section, we will approach the problem from a different viewpoint:
we assume that the disorder is weak, i.e., that plane waves are still
approximate eigenstates of the Hamiltonian governing the evolution
of photon modes, and the disorder is only a small perturbation. With
these assumptions, scattering rates can be calculated using Fermi's
golden rule. Beyond the conceptual differences between this model
(point-like scatterers) and the one that we discussed above (locally
extended scatterers), the use of Fermi's golden rule also implies
that we neglect multiple scattering events. This should not come as
a surprise, because Savona's model is exact in the sense that we calculate
the exact time evolution of the photon modes, i.e., we calculate to
infinite order in the perturbation series, while Fermi's golden rule
truncates at the first order correction. However, for short times
(short compared to the average scattering time) the differences should
not be too large.

\subsection{Unperturbed eigenstates}

Let us define the photon eigenvectors in the two cavities as 

\begin{equation}
\vert\vk_{a,b}\rangle=\varphi_{a,b}(z)e^{i\vk\cdot\vr}=\vert\varphi_{a,b}\rangle\vert\vk\rangle\,.
\end{equation}
As earlier, position vectors are understood in the plane of the cavities,
and consequently, $\vk$ is a two-dimensional wavevector. Given that
the two cavities are spatially separated, $\varphi_{a}(z),\varphi_{b}(z)$
are not only orthogonal, but they have disjoint supports. When the
disorder vanishes, the eigenstates of the \emph{coupled} system are
\begin{equation}
\vert\vk_{S,A}\rangle=\frac{1}{\sqrt{2}}\left(\vert\vk_{a}\rangle\pm\vert\vk_{b}\rangle\right)=\frac{1}{\sqrt{2}}(\vert\varphi_{a}\rangle\pm\vert\varphi_{b}\rangle)\vert\vk\rangle\,.
\end{equation}
Note that this decomposition into symmetric and anti-symmetric modes
is valid only for the case of degenerate cavities. The generalization
is straightforward, and it would not change the qualitative argument.

\subsection{Scattering matrix elements}

Using the orthogonality of states located in spatially separated cavities,
the scattering matrix element for intra-branch scattering of the symmetric
branch can be written as 

\begin{eqnarray}
M_{\vk\vkp}^{SS} & = & \langle\vk_{A}\vert V_{a}(\vr)+V_{b}(\vr)\vert\vkp_{A}\rangle\nonumber \\
 & = & \frac{1}{2}(\langle\varphi_{a}\vert+\langle\varphi_{b}\vert)\langle\vk\vert V_{a}(\vr)+V_{b}(\vr)\vert\vkp\rangle(\vert\varphi_{a}\rangle+\vert\varphi_{b}\rangle)\nonumber \\
 & = & \frac{1}{2}\langle\varphi_{a}\vert\varphi_{a}\rangle\langle\vk\vert V_{a}(\vr)\vert\vkp\rangle+\frac{1}{2}\langle\varphi_{b}\vert\varphi_{b}\rangle\langle\vk\vert V_{b}(\vr)\vert\vkp\rangle\nonumber \\
 & = & \frac{1}{2}\left[\tilde{V}_{\vk-\vkp}^{a}+\tilde{V}_{\vk-\vkp}^{b}\right]\,,\label{eq:matrix_element_ss}
\end{eqnarray}
where $\tilde{V}_{\mathbf{q}}^{a,b}$ is the Fourier transform of
the potentials. (In the matrix element $M_{\vk\vkp}^{SS}$, the first
superscript denotes the pumped mode at momentum $\vk$, while the
second superscipt is the mode to which the scattering happens at momentum
$\vkp$.) The calculation for the anti-symmetric branch is similar,
as are the results:

\begin{eqnarray}
M_{\vk\vkp}^{AA} & = & \frac{1}{2}(\langle\varphi_{a}\vert-\langle\varphi_{b}\vert)\langle\vk\vert V_{a}(\vr)+V_{b}(\vr)\vert\vkp\rangle(\vert\varphi_{a}\rangle-\vert\varphi_{b}\rangle)\nonumber \\
 & = & \frac{1}{2}\left[\tilde{V}_{\vk-\vkp}^{a}+\tilde{V}_{\vk-\vkp}^{b}\right]\,.\label{eq:matrix_element_aa}
\end{eqnarray}
From here we can draw the conclusion that the functional form of the
scattering matrix element for intra-branch scattering is independent
of whether the scattering happens on the symmetric or the anti-symmetric
state (of course, the initial and final states on the two branches
will correspond to wave vectors of different magnitude), and that
it will be zero, if the Fourier coefficients of the two potentials
are inverses of each other, i.e., when the potentials are totally
anti-correlated. Note that this is a sufficient, but not a necessary
condition. 

Also note that these are only the matrix elements, but in order to
calculate the transition probabilities, we still have to account for
energy conservation. That requirement leads to the condition $\vert\vk\vert=\vert\vkp\vert$.

Finally, we calculate the scattering matrix element for inter-branch
scattering, and we obtain

\begin{eqnarray}
M_{\vk\vkp}^{SA} & = & \langle\vk_{A}\vert V_{a}(\vr)+V_{b}(\vr)\vert\vkp_{B}\rangle\nonumber \\
 & = & \frac{1}{2}(\langle\varphi_{a}\vert+\langle\varphi_{b}\vert)\langle\vk\vert V_{a}(\vr)+V_{b}(\vr)\vert\vkp\rangle(\vert\varphi_{a}\rangle-\vert\varphi_{b}\rangle)\nonumber \\
 & = & \frac{1}{2}\left[\tilde{V}_{\vk-\vkp}^{a}-\tilde{V}_{\vk-\vkp}^{b}\right]\,.\label{eq:matrix_element_sa}
\end{eqnarray}
As for the other two cases, this expression gives only the scattering
matrix elements, but energy conservation still has to be taken into
account. We can also see that inter-branch scattering vanishes, whenever
the disorder is totally correlated in the two cavities. 

Another way of looking at this result is that, if both cavities have
the same potential, then the system still possesses reflection symmetry
with respect to the coupling mirror, and it still makes sense to talk
about symmetric and anti-symmetric wavefunctions, and in such a case,
scattering from one branch to the other is obviously forbidden by
symmetry. 

\begin{figure}[h]
\includegraphics[width=1\columnwidth]{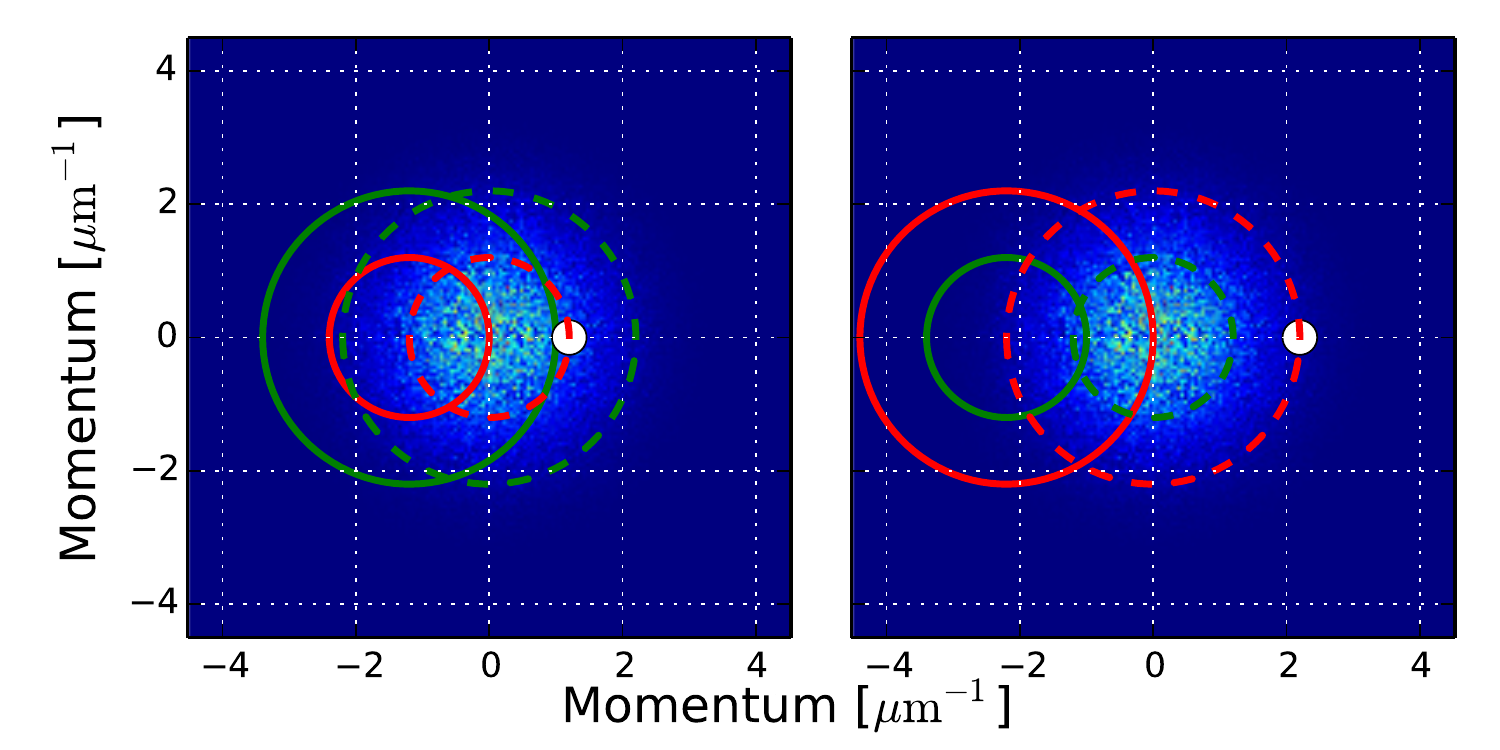}

\caption{Rayleigh scattering in momentum space in a simple coupled cavity configuration.
The white dot is the position of the pump beam ($\vk=(1.2,0)\,\mu\mathrm{m^{-1}}$
on the left hand side, $\vk=(2.2,0)\,\mu\mathrm{m^{-1}}$ on the right
hand side), while the final states are drawn in dashed lines. The
solid circles are the difference wavevectors that enter the matrix
elements in Eqs.\ (\ref{eq:matrix_element_ss}-\ref{eq:matrix_element_sa}).
In color scale, also shown is the Fourier transform of the disorder
potential with a correlation length of $2\,\mu\mathrm{m}$.}

\label{fig:coupled_scattering}
\end{figure}

It is also worth pointing out that neither for inter-branch, nor for
intra-branch scattering do the scattering rates depend explicitly
on the strength of the coupling between the two cavities. The cavity
splitting determines only the length of the momentum of the out-scattered
polaritons, and thus the Fourier components that we have to consider
for a particular pump angle. The scattering wavevectors are displayed
in Fig.\,\ref{fig:coupled_scattering}. The two cavities are pumped
on either the symmetric branch ($S$) at at $\vk=(2.2,0)\,\mu\mathrm{m^{-1}}$,
or the anti-symmetric branch ($A$) at $\vk=(1.2,0)\,\mu\mathrm{m^{-1}}$
(white dot), and scattering can happen to the wavevectors drawn by
the dashed circles (red is the pumped, while green is the unpumped
branch). These are the states whose energy is the same as that at
the pump momentum. The difference wavevectors are depicted by the
solid lines, and in order to find the scattering intensity on a particular
branch, one has to integrate over one of these two circles.

\subsection{Scattering intensities}

At this point, we can write the potentials as in Eqs.\,(\ref{eq:va_numerical}-\ref{eq:vb_numerical}),
and get for the Fourier transforms

\begin{equation}
\tilde{V}_{\vk}^{a,b}=V_{0}\tilde{K}_{\vk}(\xi)\tilde{N}_{\vk}^{a,b}\,,
\end{equation}
where $\tilde{K}_{\vk}(\xi)$, and $\tilde{N_{\vk}}^{a,b}$ are the
Fourier transforms of the Gaussian kernel, and the random potential,
respectively. Hence, the total scattering intensities from branch
$i$ to branch $j$ are 

\begin{widetext}
\begin{eqnarray}
I^{ij} & \sim & V_{0}^{2}\int\, d\vk\left|\tilde{V}_{\vk}^{a}\pm\tilde{V}_{\vk}^{b}\right|^{2}=V_{0}^{2}\int\, d\vk\tilde{K}_{\vk}^{2}(\xi)\left|(1\pm\mathcal{C})\tilde{N}_{\vk}^{a}\pm\sqrt{1-\mathcal{C}^{2}}\tilde{N}_{\vk}^{b}\right|^{2}\nonumber \\
 & = & V_{0}^{2}\int\, d\vk\tilde{K}_{\vk}^{2}(\xi)\left\{ (1\pm\mathcal{C})^{2}\left|\tilde{N}_{\vk}^{a}\right|^{2}+\left(1-\mathcal{C}^{2}\right)\left|\tilde{N}_{\vk}^{a}\right|^{2}\pm(1\pm\mathcal{C})\sqrt{1-\mathcal{C}^{2}}\left[\tilde{N}_{\vk}^{a}(\tilde{N}_{\vk}^{b})^{*}+(\tilde{N}_{\vk}^{a})^{*}\tilde{N}_{\vk}^{b}\right]\right\} \nonumber \\
 & = & V_{0}^{2}\int\, d\vk\tilde{K}_{\vk}^{2}(\xi)\left\{ (1\pm\mathcal{C})^{2}\left|\tilde{N}_{\vk}^{a}\right|^{2}+\left(1-\mathcal{C}^{2}\right)\left|\tilde{N}_{\vk}^{a}\right|^{2}\right\} \nonumber \\
 & = & 2(1\pm\mathcal{C})V_{0}^{2}\int\, d\vk\tilde{K}_{\vk}^{2}(\xi)\left|\tilde{N}_{\vk}^{a}\right|^{2}=2(1\pm\mathcal{C})V_{0}^{2}\mathcal{F}_{\pm}^{ij}\left(\left|\vk_{\mathrm{in}}^{i}\right|,\left|\vk_{\mathrm{out}}^{j}\right|,\xi\right)
\end{eqnarray}

\end{widetext}because the cross-terms in the second line drop out,
given that the two random variables are not correlated. We should
point out that the intensity can be written in this simple form only
when the disorder potential can be described by the same Gaussian
kernel in the two cavities, i.e., when the transverse correlation
lengths are identical. As in the numerical model, the integration
is understood to be on a contour allowed by energy conservation, and
the last line defines the function $\mathcal{F}_{\pm}^{ij}$, which
measures the scattering amplitude. $\hbar\left|\vk_{\mathrm{in}}^{i}\right|$,
and $\hbar\left|\vk_{\mathrm{out}}^{j}\right|$ are the length of
the in-coming and out-scattered momenta. Therefore, the visibilities
defined in Eq.\,(\ref{eq:visibilities}) become

\begin{equation}
v_{S}=\frac{\left(\mathcal{F}_{+}^{SS}-\mathcal{F}_{-}^{SA}\right)+\mathcal{C}\left(\mathcal{F}_{+}^{SS}+\mathcal{F}_{-}^{SA}\right)}{\left(\mathcal{F}_{+}^{SS}+\mathcal{F}_{-}^{SA}\right)+\mathcal{C}\left(\mathcal{F}_{+}^{SS}-\mathcal{F}_{-}^{SA}\right)}=\frac{\gamma+\mathcal{C}}{1+\gamma\mathcal{C}}\,,\label{eq:vs_analytical}
\end{equation}
and

\begin{equation}
v_{A}=\frac{\left(\mathcal{F}_{+}^{AA}-\mathcal{F}_{-}^{AS}\right)+\mathcal{C}\left(\mathcal{F}_{+}^{AA}+\mathcal{F}_{-}^{AS}\right)}{\left(\mathcal{F}_{+}^{AA}+\mathcal{F}_{-}^{AS}\right)+\mathcal{C}\left(\mathcal{F}_{+}^{AA}-\mathcal{F}_{-}^{AS}\right)}=\frac{\mathcal{C}-\gamma'}{1-\gamma'\mathcal{C}}\,,\label{eq:va_analytical}
\end{equation}
respectively. Here, we introduced the abbreviation 

\[
\gamma=\frac{\mathcal{F}_{+}^{SS}-\mathcal{F}_{-}^{SA}}{\mathcal{F}_{+}^{SS}+\mathcal{F}_{-}^{SA}}\,,\quad\mathrm{and}\quad\gamma'=\frac{\mathcal{F}_{+}^{AA}-\mathcal{F}_{-}^{AS}}{\mathcal{F}_{+}^{AA}+\mathcal{F}_{-}^{AS}}\,.
\]

\begin{figure}[h]
\includegraphics[width=1\columnwidth]{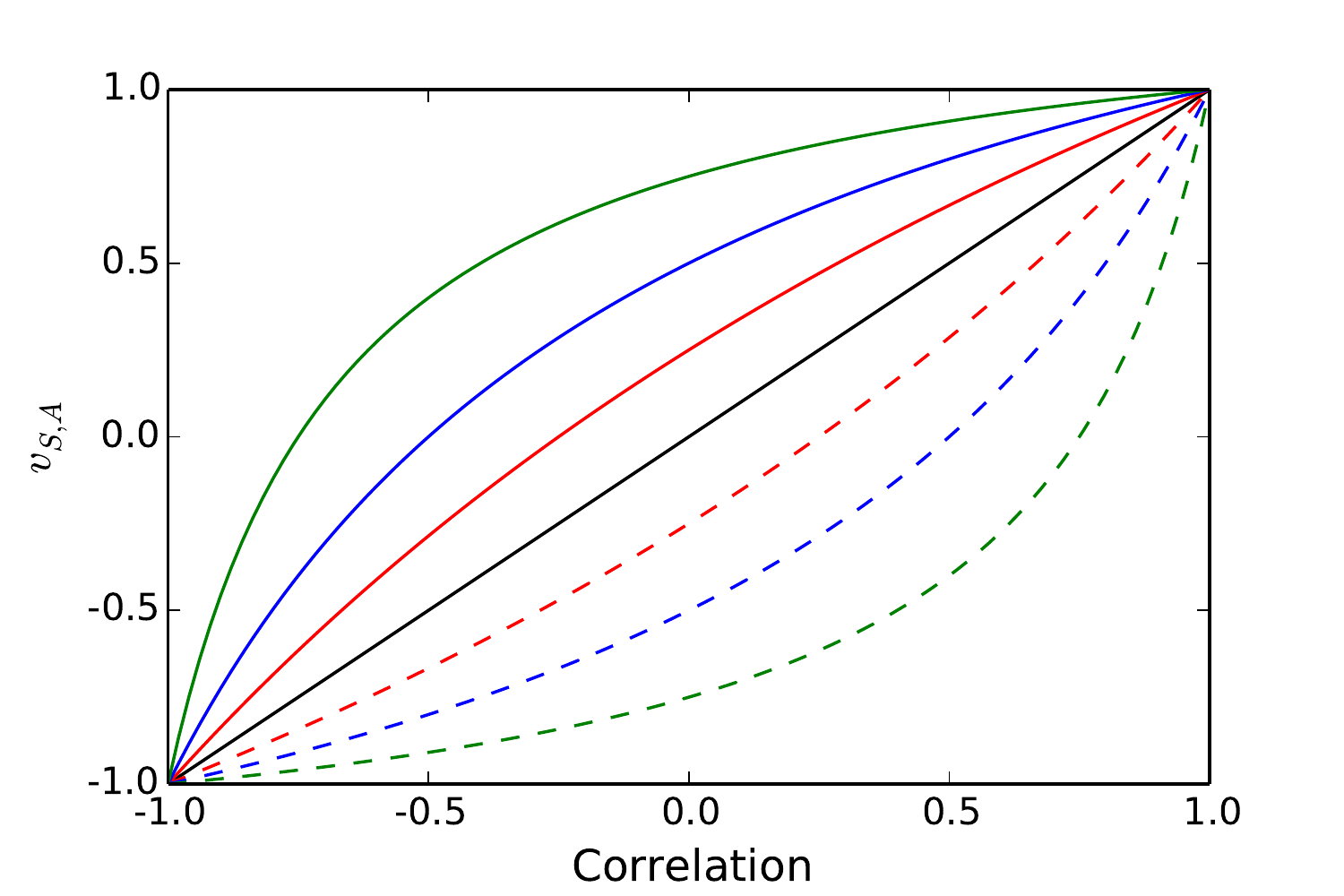}

\caption{The curves $v_{S,A}$ as calculated from Eqs.\,(\ref{eq:vs_analytical}-\ref{eq:va_analytical})
for $\gamma=0$ (black), $\gamma=1/4$ (red), $\gamma=1/2$ (blue),
and $\gamma=3/4$ (green). The solid lines correspond to $v_{S}$
(Eq.\,(\ref{eq:vs_analytical})), while the dashed lines show $v_{A}$
(Eq.\,(\ref{eq:va_analytical})).}

\label{fig:vas_analytical}
\end{figure}

We can make some general remarks even without explicitly calculating
the scattering intensities. $v_{S,A}(\mathcal{C}=\pm1)=\pm1$, independent
of the value of $\gamma,\gamma'$ which only determine how the two
endpoints are connected. Since, by definition, $\gamma,\gamma'<1$,
the curves behave in a strictly monotonic fashion as a function of
$\mathcal{C}$. Also note that the two curves corresponding to the
same value of $\gamma$, and $\gamma'$ possess inversion symmetry
around the origin, as can also be seen in Fig.\,(\ref{fig:vas_analytical}).

\subsection{Special case: Gaussian correlations}

There are further simplifications that we can make: since $N(\vr)$
is a random variable, so is $\tilde{N}_{\vk}$, and we can then replace
$\vert\tilde{N}_{\vk}\vert^{2}$ by its average. If we do that, however,
then $\mathcal{F}_{\pm}^{ij}$ becomes a contour integral over a Gaussian
surface (the solid lines in Fig.\,\ref{fig:coupled_scattering}),
and we obtain

\begin{eqnarray}
I^{ij} & \sim & \int e^{-\vert\vk^{i}-\vkp^{j}\vert^{2}\xi^{2}}\nonumber \\
{} & = & e^{-(k_{\mathrm{out}}^{2}+k_{\mathrm{in}}^{2})\xi^{2}}\int\, d\alpha e^{2k_{\mathrm{out}}k_{\mathrm{in}}\cos(\alpha)\xi^{2}}\,,\label{eq:simplified_intensity}
\end{eqnarray}
where $\alpha$ is the angle between the in-coming and out-going wavevectors.
In the special cases, when the integration is over the $[0,2\pi]$
domain we get the modified Bessel function $I_{0}(2k_{\mathrm{out}}k_{\mathrm{in}}\xi^{2})$,
and when it is over the $[\pi/2,3\pi/2]$ domain, the modified Struve
function $L_{0}(2k_{\mathrm{out}}k_{\mathrm{in}}\xi^{2})$. It is
also clear from the particular form in Eq.\,(\ref{eq:simplified_intensity})
that the relevant momentum scale is the inverse of the transverse
correlation length. 

\begin{figure}[h]
\includegraphics[width=1\columnwidth]{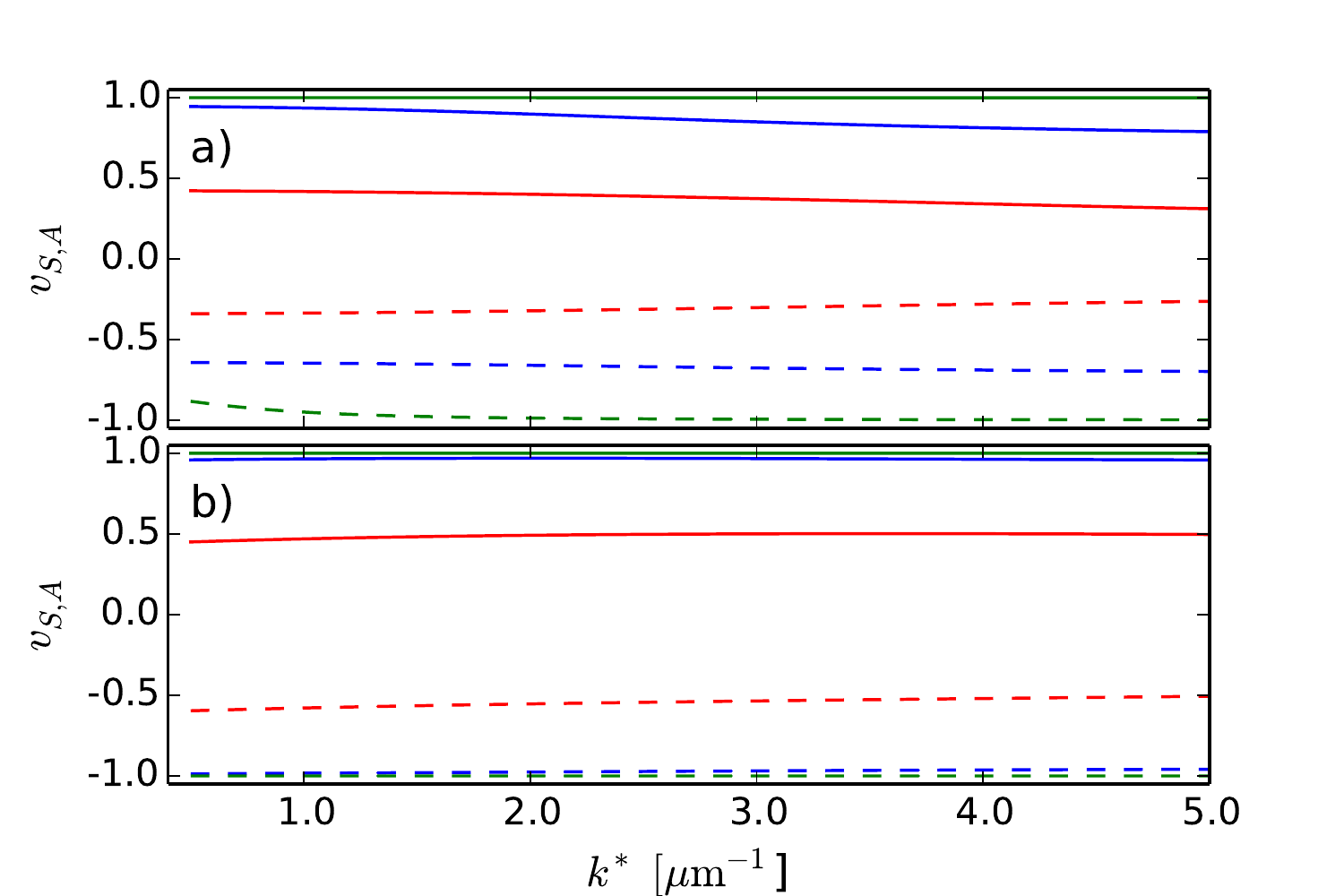}

\caption{The dependence of the scattering visibility on the excitation momentum
for three values of the transverse correlation length: $\xi=0.25$
(red) $0.5$ (blue), and $1\,\mu\mathrm{m}$ (green). The solid lines
show $v_{S}$, while the dashed lines are $v_{A}$. The transverse
cross-correlation was set to $\mathcal{C}=0$. The plot on the top
shows curves for an integration domain of $[0,2\pi]$, and on the
bottom $[\pi/2,3\pi/2]$.}

\label{fig:visibility_vs_momentum}
\end{figure}

Experimentally, it is probably hard to grow a sample with a prescribed
transverse cross-correlation, and these correlations will also depend
on the exact position on the sample. The transverse cross-correlation,
therefore, is not a convenient control parameter. It does not mean,
however, that the model cannot be tested. First, the very spatial
variation can be exploited to realize different values of $\mathcal{C}$.
Second, one can calculate the momentum dependence of the visibility
at a fixed position, therefore, fixed correlation. The excitation
energy fixes both the input and output wavevectors, and they can be
calculated from the dispersion relations. In Fig.\,\ref{fig:visibility_vs_momentum}
we plot the visibility as a function of the excitation momentum for
the experimentally realistic case of 14\,meV mode splitting and cavity
length $L=800$\,nm, and for three different values of the transverse
correlation length, $\xi$, with a transverse cross-correlation of
$\mathcal{C}=0$. The integration domain was $[0,2\pi]$ (a), and
$[\pi/2,3\pi/2]$ (b). Keeping in mind that in a reflection configuration
the directly reflected light has to be blocked, the second domain
is a more plausible experimental condition. The horizontal axis is
always the shorter wavevector, i.e., $k^{*}=k_{\mathrm{in}}^{S}$
(pumping is on the symmetric branch), or $k^{*}=k_{\mathrm{out}}^{S}$
(pumping is on the anti-symmetric branch).

We can observe that independent of the transverse correlation length,
the visibilities do not show a strong momentum dependence for case
b). The physical reason for this is that (c.f. Fig.\,\ref{fig:coupled_scattering})
the integration domain $[\pi/2,3\pi/2]$ samples the Gaussian kernel
far from the center, where the function is relatively flat, but still
decreasing. Therefore, by increasing the momentum, we integrate a
slightly smaller integrand along a larger half circle, with the net
result of obtaining approximately the same number. One would expect
the same flat dependence from a simple filter theory, therefore, this
measurement alone would not be able to distinguish between the various
theories.

We can also notice that larger correlation lengths lead to larger
absolute values for the visibilities. The interpretation of this is
that for larger correlation lengths, the Gaussian kernel in momentum
space is smaller, resulting in a more imbalanced intensity distribution
on the two scattering circles. However, as the imbalance increases,
so does the visibility. 

As seen in Fig.\,\ref{fig:visibility_vs_momentum}a), when we integrate
over the domain $[0,2\pi]$, the visibilities depend more strongly
on the momentum. This can be understood, if we consider that in this
case, the integration contour samples not only the flat, but the rapidly
varying parts of the kernel, and it is no longer true that an increase
in the length of the integration contour would be off set by the drop
in the value of the kernel. 

\section{Conclusion}

In conclusion, we have developed a theory for resonant Rayleigh scattering
in coupled planar microcavity structures. In the first part of the
paper, we extended Savona's theory to the case of coupling, and presented
some numerical results. This theory solves the exact time evolution
of the photonic component of polariton scattering, and as such, is
exact for all times. In the second part of the paper, we developed
an analytical approach, and derived results. Given the perturbative
nature of this model, these results can be trusted at short times
only. 

The main result of our paper is that depending on the transverse cross-correlations
of the disorder potentials, RRS can be either enhanced, or suppressed
on a particular polariton branch. Keeping an eye on parametric scattering
schemes aiming to produce quantum correlated light, this result also
means that it might be beneficial to use coupled cavities, even in
the cases, where only a single polariton branch is involved, because
if one can find a position on the sample, where the transverse cross-correlation
is negative, then the relative strength of the parametric signal could
be boosted on the branch under consideration.
\begin{acknowledgments}
The authors are indebted to Sebastian Diehl and Vincenzo Savona for
the illuminating discussions. We also acknowledge financial support
from the Austrian Science Fund, FWF, project number P-22979-N16.
\end{acknowledgments}
\bibliographystyle{apsrev}
\bibliography{polariton}

\appendix

\section{Scattering matrix elements in a triple cavity}

\begin{widetext}Triple microcavities are used frequently in vertical
(pump, signal and idler all are at zero momentum) parametric oscillators
\cite{Diederichs2005,Diederichs2006,Diederichs2007,Diederichs2007a,Diederichs2007b,Taj2009},
or, as proposed in Portolan et al. \cite{Portolan2013}, as a source
of hyper-entangled photons. For this reason, we briefly discuss how
the analytical model can be written down for such a case. When three
identical cavities are pairwise coupled with equal couplings, the
new eigenstates can be written as two completely symmetric, and a
completely anti-symmetric linear combination of the original states:

\begin{eqnarray*}
\vert\vk_{A}\rangle & = & \frac{1}{\sqrt{2}}(\vert\vk_{a}\rangle-\vert\vk_{c}\rangle)\\
\vert\vk_{B}\rangle & = & \frac{1}{2}\vert\vk_{a}\rangle-\frac{1}{\sqrt{2}}\vert\vk_{b}\rangle+\frac{1}{2}\vert\vk_{c}\rangle\\
\vert\vk_{C}\rangle & = & \frac{1}{2}\vert\vk_{a}\rangle+\frac{1}{\sqrt{2}}\vert\vk_{b}\rangle+\frac{1}{2}\vert\vk_{c}\rangle
\end{eqnarray*}
The scattering matrix element can be calculated as in Eqs.\,(\ref{eq:matrix_element_ss}-\ref{eq:matrix_element_sa}),
and we obtain the following results

\begin{eqnarray*}
M_{\vk\vkp}^{AA} & = & \langle\vk_{A}\vert V_{a}+V_{b}+V_{c}\vert\vkp_{A}\rangle=\frac{1}{2}(\tilde{V}_{\vk-\vkp}^{a}+\tilde{V}_{\vk-\vkp}^{c})\\
M_{\vk\vkp}^{AB} & = & \langle\vk_{A}\vert V_{a}+V_{b}+V_{c}\vert\vkp_{B}\rangle=\frac{1}{2\sqrt{2}}(\tilde{V}_{\vk-\vkp}^{a}-\tilde{V}_{\vk-\vkp}^{c})\\
M_{\vk\vkp}^{AC} & = & \langle\vk_{A}\vert V_{a}+V_{b}+V_{c}\vert\vkp_{C}\rangle=\frac{1}{2\sqrt{2}}(\tilde{V}_{\vk-\vkp}^{a}+\tilde{V}_{\vk-\vkp}^{c})\\
M_{\vk\vkp}^{BB} & = & \langle\vk_{B}\vert V_{a}+V_{b}+V_{c}\vert\vkp_{B}\rangle=\frac{1}{4}\tilde{V}_{\vk-\vkp}^{a}+\frac{1}{2}\tilde{V}_{\vk-\vkp}^{b}+\frac{1}{4}\tilde{V}_{\vk-\vkp}^{c}\\
M_{\vk\vkp}^{BC} & = & \langle\vk_{B}\vert V_{a}+V_{b}+V_{c}\vert\vkp_{C}\rangle=\frac{1}{4}\tilde{V}_{\vk-\vkp}^{a}-\frac{1}{2}\tilde{V}_{\vk-\vkp}^{b}+\frac{1}{4}\tilde{V}_{\vk-\vkp}^{c}\\
M_{\vk\vkp}^{CC} & = & \langle\vk_{C}\vert V_{a}+V_{b}+V_{c}\vert\vkp_{C}\rangle=\frac{1}{4}\tilde{V}_{\vk-\vkp}^{a}+\frac{1}{2}\tilde{V}_{\vk-\vkp}^{b}+\frac{1}{4}\tilde{V}_{\vk-\vkp}^{c}
\end{eqnarray*}

The potentials will be written as in Eqs.\,(\ref{eq:va_numerical}-\ref{eq:vb_numerical}):

\begin{eqnarray}
V_{a}(\vr) & = & V_{0}\int\, d\vrp K(\vr,\vrp,\xi)N_{a}(\vrp)\label{eq:app_va_numerical}\\
V_{b}(\vr) & = & \mathcal{C}_{a}V_{a}(\vr)+V_{0}\sqrt{1-\mathcal{C}_{a}^{2}}\int\, d\vrp K(\vr,\vrp,\xi)N_{b}(\vrp)\label{eq:app_vb_numerical}\\
V_{c}(\vr) & = & \mathcal{C}_{b}V_{b}(\vr)+V_{0}\sqrt{1-\mathcal{C}_{b}^{2}}\int\, d\vrp K(\vr,\vrp,\xi)N_{c}(\vrp)\label{eq:app_vc_numerical}
\end{eqnarray}

or, in terms of the Fourier components, 

\begin{eqnarray*}
\tilde{V}_{\vk}^{a} & = & V_{0}\tilde{K}_{\vk}(\xi)\tilde{N}_{\vk}^{a}\\
\tilde{V}_{\vk}^{b} & = & V_{0}\tilde{K}_{\vk}(\xi)(\mathcal{C}_{a}\tilde{N}_{\vk}^{a}+\sqrt{1-\mathcal{C}_{a}^{2}}\tilde{N}_{\vk}^{b})\\
\tilde{V}_{\vk}^{c} & = & V_{0}\tilde{K}_{\vk}(\xi)(\mathcal{C}_{a}\mathcal{C}_{b}\tilde{N}_{\vk}^{a}+\mathcal{C}_{b}\sqrt{1-\mathcal{C}_{a}^{2}}\tilde{N}_{\vk}^{b}+\sqrt{1-\mathcal{C}_{b}^{2}}\tilde{N}_{\vk}^{c})
\end{eqnarray*}

and hence, the total scattered intensities can be expressed as 

\begin{eqnarray*}
I^{AA} & = & \int\, d\vk\left|M_{\vk\vkp}^{AA}\right|{}^{2}=\frac{V_{0}^{2}}{4}\int\, d\vk\tilde{K}_{\vk}^{2}(\xi)\left|\tilde{V}_{\vk}^{a}+\tilde{V}_{\vk}^{c}\right|^{2}\\
 & = & \frac{V_{0}^{2}}{4}\int\, d\vk\tilde{K}_{\vk}^{2}(\xi)\left\{ (1+\mathcal{C}_{a}\mathcal{C}_{b})^{2}\left|\tilde{N}_{\vk}^{a}\right|^{2}+\mathcal{C}_{b}^{2}(1-\mathcal{C}_{a}^{2})\left|\tilde{N}_{\vk}^{b}\right|^{2}+(1-\mathcal{C}_{b}^{2})\left|\tilde{N}_{\vk}^{c}\right|^{2}\right\} \\
 & = & \frac{V_{0}^{2}}{4}\left[(1+\mathcal{C}_{a}\mathcal{C}_{b})^{2}+\mathcal{C}_{b}^{2}(1-\mathcal{C}_{a}^{2})+(1-\mathcal{C}_{b}^{2})\right]\int\, d\vk\tilde{K}_{\vk}^{2}(\xi)\left|\tilde{N}_{\vk}^{a}\right|^{2}\\
 & = & \frac{V_{0}^{2}}{4}\left[(1+\mathcal{C}_{a}\mathcal{C}_{b})^{2}+\mathcal{C}_{b}^{2}(1-\mathcal{C}_{a}^{2})+(1-\mathcal{C}_{b}^{2})\right]\mathcal{F}\left(\left|\vk_{\mathrm{in}}^{A}\right|,\left|\vk_{\mathrm{out}}^{A}\right|,\xi\right)\\
 & = & \frac{V_{0}^{2}}{2}(1+\mathcal{C}_{a}\mathcal{C}_{b})\mathcal{F}\left(\left|\vk_{\mathrm{in}}^{A}\right|,\left|\vk_{\mathrm{out}}^{A}\right|,\xi\right)
\end{eqnarray*}
because, due to the statistical independence of the noise distributions
$\tilde{N}_{\vk}^{a,b,c}$, all cross-terms vanish upon integration
over a finite domain. All other intensites can be calculated in the
same fashion, and we get 

\begin{eqnarray*}
I^{AB} & = & \frac{V_{0}^{2}}{4}(1-\mathcal{C}_{a}\mathcal{C}_{b})\mathcal{F}\left(\left|\vk_{\mathrm{in}}^{A}\right|,\left|\vk_{\mathrm{out}}^{B}\right|,\xi\right)\\
I^{AC} & = & \frac{V_{0}^{2}}{4}(1+\mathcal{C}_{a}\mathcal{C}_{b})\mathcal{F}\left(\left|\vk_{\mathrm{in}}^{A}\right|,\left|\vk_{\mathrm{out}}^{C}\right|,\xi\right)\\
I^{BB} & = & \frac{V_{0}^{2}}{16}\left(6+\mathcal{C}_{a}^{2}\mathcal{C}_{b}^{2}+4\mathcal{C}_{b}\sqrt{1-\mathcal{C}_{a}^{2}}+2\mathcal{C}_{a}\left(2+\mathcal{C}_{b}+2\mathcal{C}_{a}\mathcal{C}_{b}\right)\right)\mathcal{F}\left(\left|\vk_{\mathrm{in}}^{B}\right|,\left|\vk_{\mathrm{out}}^{B}\right|,\xi\right)\\
I^{BC} & = & \frac{V_{0}^{2}}{16}\left(6+\mathcal{C}_{a}^{2}\mathcal{C}_{b}^{2}-4\mathcal{C}_{b}\sqrt{1-\mathcal{C}_{a}^{2}}-2\mathcal{C}_{a}\left(2-\mathcal{C}_{b}+2\mathcal{C}_{a}\mathcal{C}_{b}\right)\right)\mathcal{F}\left(\left|\vk_{\mathrm{in}}^{B}\right|,\left|\vk_{\mathrm{out}}^{C}\right|,\xi\right)\\
I^{CC} & = & \frac{V_{0}^{2}}{16}\left(6+\mathcal{C}_{a}^{2}\mathcal{C}_{b}^{2}+4\mathcal{C}_{b}\sqrt{1-\mathcal{C}_{a}^{2}}+2\mathcal{C}_{a}\left(2+\mathcal{C}_{b}+2\mathcal{C}_{a}\mathcal{C}_{b}\right)\right)\mathcal{F}\left(\left|\vk_{\mathrm{in}}^{C}\right|,\left|\vk_{\mathrm{out}}^{C}\right|,\xi\right)
\end{eqnarray*}
The expressions for $I^{BA}$,$I^{CA}$, and $I^{CB}$ can be obtained
by reversing the order of the arguments in $\mathcal{F}$:

\begin{eqnarray*}
I^{BA} & = & \frac{V_{0}^{2}}{4}(1-\mathcal{C}_{a}\mathcal{C}_{b})\mathcal{F}\left(\left|\vk_{\mathrm{in}}^{B}\right|,\left|\vk_{\mathrm{out}}^{A}\right|,\xi\right)\\
I^{CA} & = & \frac{V_{0}^{2}}{4}(1+\mathcal{C}_{a}\mathcal{C}_{b})\mathcal{F}\left(\left|\vk_{\mathrm{in}}^{C}\right|,\left|\vk_{\mathrm{out}}^{A}\right|,\xi\right)\\
I^{CB} & = & \frac{V_{0}^{2}}{16}\left(6+\mathcal{C}_{a}^{2}\mathcal{C}_{b}^{2}-4\mathcal{C}_{b}\sqrt{1-\mathcal{C}_{a}^{2}}-2\mathcal{C}_{a}\left(2-\mathcal{C}_{b}+2\mathcal{C}_{a}\mathcal{C}_{b}\right)\right)\mathcal{F}\left(\left|\vk_{\mathrm{in}}^{C}\right|,\left|\vk_{\mathrm{out}}^{B}\right|,\xi\right)
\end{eqnarray*}

\end{widetext}
\end{document}